\documentstyle[prl,aps,psfig,twocolumn]{revtex} 

\begin{document}
\draft
\twocolumn[\hsize\textwidth\columnwidth\hsize\csname
@twocolumnfalse\endcsname
\title{Putting Proteins back into Water.}
\author{Paolo De Los Rios$^1$, Guido Caldarelli$^2$}
\address{$^1$Institut de Physique Th\'eorique, 
Universit\'e de Fribourg, CH-1700, Fribourg, Switzerland.}
\address{$^2$INFM Sezione di Roma1, Dip. Fisica Universit\`a La Sapienza, 
Ple A. Moro 2, 00185 Roma Italy.}
\date{\today}
\maketitle

\begin{abstract}
We introduce a simplified protein model
where the solvent (water) degrees of freedom
appear explicitly (although in an extremely simplified fashion).
Using this model we are able to recover the thermodynamic
phenomenology of proteins over a wide range of temperatures. 
In particular we describe both the warm and the {\it cold}
protein denaturation within a single framework, while addressing
important issues about the structure of model proteins. 
\end{abstract}
]
\narrowtext

One of the main goals of statistical physics in the last decade has been to 
understand the "folding code": how the amino-acid sequence
of a protein (coded in DNA, the "genetic code"),
uniquely determines its functional ("native") structure, or fold.
Understanding the principles that drive a protein to fold to its native
structure is of great conceptual and practical relevance, since it could lead,
for example, to high specificity drugs. 

Proteins are extremely complex structures: they are long heteropolymers 
made of up to $20$ different amino-acids species, 
each of them with its own chemical, 
electrostatic and steric properties; the physiological solvent, an aqueous
solution, and its characteristics play a fundamental role both in the 
dynamics and in the thermodynamics of folding. 
It is therefore not surprising that only in recent times 
statistical physicists have begun working
on this problem, mainly after the introduction of the so-called HP
model\cite{LD89+},
where the above mentioned richness has been reduced to a manageable 
level. 
In the HP model, proteins are modeled as self-avoiding polymers
on a lattice (two or three dimensional), greatly reducing
the number of accessible conformations\cite{Vanderzande}. 
The chemical and electrostatic properties 
of amino-acids have also been simplified: indeed, 
it has been recognized since long 
that the main force stabilizing the native conformations of 
globular proteins is the {\it hydrophobicity} of non-polar
amino-acids\cite{Kauzmann59}.
Consequently, the important
properties of amino-acids are reduced to two: they are either polar
(ions or dipoles, labeled with P) or non-polar (H).

Hydrophobicity can be described as the tendency of hydrophobic molecules
to reduce as much as possible their surface of contact with water:
two hydrophobic molecules try to stick together in order 
to hide from water their mutual surface of contact. Consequently, 
hydrophobicity has been introduced in the HP  model as an effective 
attractive interaction between H amino-acids. Then, the solvent
degrees of freedom can be neglected. Here we show that such a simplification 
can be removed, and water can be taken into account, keeping the complexity 
of the model at a still manageable level: the benefits are a 
better description of the protein phenomenology (namely, {\it cold} 
destabilization and eventually denaturation\cite{MPG90,Privalov1}) 
and some insights on the structure of the protein core. 

In the last fifteen years there has been a growing body of evidence
for the so called {\it cold destabilisation} of proteins: the free energy
difference $\Delta F^N_D$ between denaturate and native
conformations of proteins has parabolic shape, with a maximum
at temperatures of the order of $15-25 ^0$C, or lower, implying that at lower 
temperatures the native conformation is less and less stable. 
In some cases, even the {\it cold denaturation}
of proteins has been obtained~\cite{Privalov2}.

The HP model is unable to deal with cold destabilization since its 
low temperature state is compact and more and more stable down to $T=0$:
is a good description of cold destabilisation and
eventually denaturation relevant for protein folding?
We think that the answer is affirmative for at least two reasons.
 
In order to describe protein folding with a simple model, it is
important to capture the essential physics of the process, at the temperatures
at which it takes place.
If the stability of native conformations of proteins begins to decrease
below $15-25 ^o$C, it is unlikely, at least {\it a priori},
that the physics responsible for such a behavior is not important around
the maximal stability temperature, in a range relevant for {\it in vivo} 
protein folding.
A further reason to believe that a good model for protein folding should also
agree with the cold destabilisation phenomenology is that, actually,
there is no clear-cut distinction between the physics that stabilises proteins,
and the one that destabilises them. In both cases a re-analysis
of the concept of hydrophobicity and of hydrophobic hydration is necessary.

Already Frank and Evans~\cite{FE45+} identified the origin of hydrophobicity in 
the partial ordering of water around non-polar molecules (such as, for
example, pentane, benzene and some amino-acids). 
Water molecules tend to build ice-like cages around non-polar molecules. 
Although a detailed analysis of these structures is,
to our knowledge, still lacking (actually recently some
better understanding and consensus are emerging\cite{Muller90,LG96,SHD98,SHD99}),
we can guess their energetic and entropic properties. Indeed,
water molecules forming these cages are highly hydrogen bonded, much as
in ice; consequently, their formation is energetically favorable with respect
to bulk liquid water.
Yet, the possible molecular arrangements in the cages are a small number
compared to all the disordered molecular conformations typical of liquid water.
The latter are
energetically unfavorable with respect to bulk water because
water molecules fail to form hydrogen bonds with hydrophobic amino-acids.
Therefore the free energy of formation of a cage ($F_{cage}-F_{no\;cage} =
\Delta F$) is a 
balance between an enthalpy gain/loss and an entropy loss/gain:
ordered cages give an enthalpy gain ($\Delta H < 0$) and an entropy loss 
($\Delta S <0$); the scenario is the opposite for disordered states.
All of the above arguments call for a model able to reproduce 
(at least qualitatively) such a rich phenomenology.

\begin{figure}
\centerline{\psfig{file=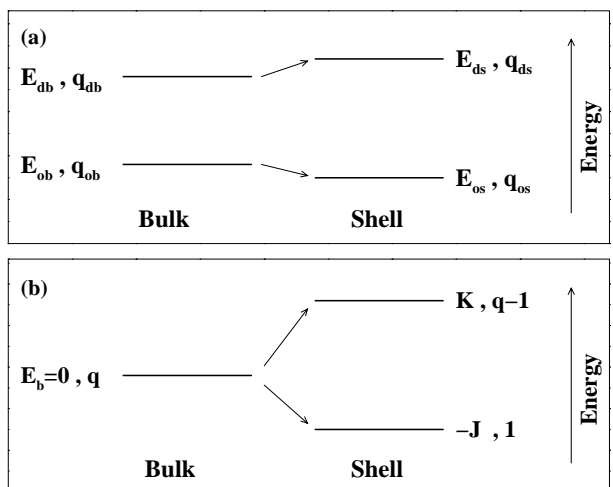,width=8.0cm}}
\caption{Bimodal effective models. Panel (a): MLG model, with bimodal energy
distributions both for bulk and shell water molecules. The lower levels
represent ordered group of water molecules, the higher levels disordered ones.
The order of energies and of degeneracies, as obtained from experiments,
is $E_{ds} > E_{db} > E_{ob} > E_{os}$ and $q_{ds} > q_{db} > q_{ob} > q_{os}$
($ds$ = disordered shell, $os$ = ordered shell, $db$ = disordered bulk, $ob$ = ordered
bulk). Panel (b): the simplified bimodal energy distribution, with just two 
free parameters, $K$ and $q$, since we can take $J$ as energy scale.}
\label{Fig1}
\end{figure}

The model we propose here borrows two of the simplifications from the HP
model: proteins are still modeled as heteropolimers on a lattice, 
made of just two different amino-acid species: polar (P) and non-polar (H).
Then, we {\it put proteins back into water}: 
every site of the lattice that is not occupied by the
polymer is occupied by water (in general, by a group of water molecules that can
be arranged in $q$ states). Water is described using the
Muller-Lee-Graziano (MLG) two-states model (Fig.\ref{Fig1}a)~\cite{Muller90,LG96}, 
that Silverstein {\it et al.} have recently shown to be consistent
with a molecular model of the water-amino acid system~\cite{SHD99}. 
The energy of each $H$ amino-acid depends on the states
of the water sites it is in contact with: as a simplifying assumption (see Fig.
\ref{Fig1}b), we say
that out of the $q$ possible states of a water site, one can be singled out
to be a cage conformations (labeled $s=0$), 
energetically favorable with energy $-J$
($J>0$), and the remaining $q-1$ ($s=1,...,q-1$) 
states are energetically unfavorable with energy $K>0$
(they represent the disordered states of reduced hydrogen-bond coordination). 
We stress that the term {\it (un)favorable}
is always with respect to bulk liquid water. Water sites that are not 
in contact with $H$ amino-acids (that is, {\it bulk} water sites) 
do not contribute to the energy (whereas they would have an energetic 
description according to the MLG model, that yet has five free parameters, 
too many for a simple theoretical model).
$P$ amino-acids do not
interact with water so that their energy is always $0$: such a crude
approximation is made with the idea that hydrophobicity is the leading
effect stabilizing the native conformation of proteins. Some 
better description of the water-P 
interaction would be welcome, but
such ingredient is unnecessary for our present purposes. 

Given a protein of $N$ amino-acids, with the sequence 
$a_1,a_2,...,a_N$ ($a_i=P$ or $H$), the energy of
the protein is then
\begin{equation}
E = \sum_{<i,H>}{}(-J \delta_{s_j,0}+K(1-\delta_{s_j,0}))\;
\label{hamiltonian}
\end{equation}
where the sum is over the water sites that are nearest neighbors of some $H$
amino-acid.
Starting from (\ref{hamiltonian}) we can write the partition
function of the system as 
\begin{equation}
Z_N = \sum_{C} Z_N(C)
\label{partition}
\end{equation}
where $Z_N(C)$ is the partition function associated to a single
conformation $C$:
\begin{equation}
Z_N(C) = q^{n_0(C)}\left((q-1) e^{-\beta K } + e^{\beta J} \right)^{n_1(C)}
\label{conf part func}
\end{equation}
where the dependence on the water degrees of freedom has been explicitly 
calculated. $n_1(C)$ is the number of water sites nearest neighbors of
some $H$ amino-acid, $n_0$ is the number of bulk water sites. 

We deal with model proteins of length up to $N=17$ on the square lattice, and 
compute the partition function, and all the thermodynamic quantities and
averages
by exact enumeration of the 
$2155667$
different conformations.
We show the results for the particular sequence 
$PHPPHPPHPHPPHPPHH$.
We choose $J=1$ (actually, 
both $K$ and the temperature $T$ can be normalized with respect to $J$),
$K=2$ and $q=10^5$ (a better determination of these values
could come from molecular dynamics and structural studies).
We take the Boltzmann constant $k_B=1$.

\begin{figure}
\centerline{\psfig{file=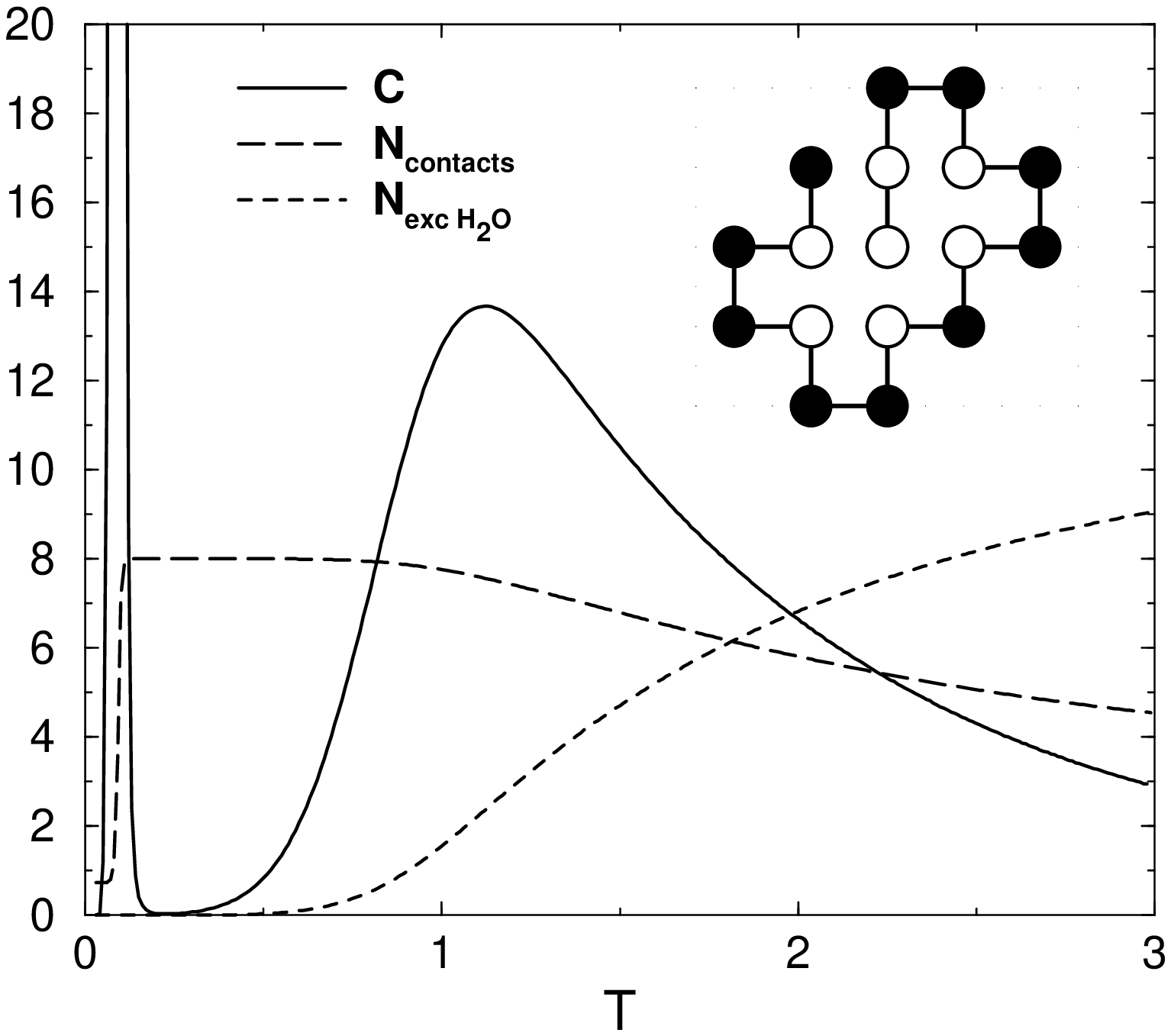,width=7.5cm}}
\caption{Specific heat, monomer-monomer contacts and number
of water sites in an excited state for the protein shown in the inset; 
$J=1$, $K=2$ and $q=10^5$.}
\label{Fig2}
\end{figure}

In Fig.\ref{Fig2} the specific heat $C_v$, and the average number of
monomer-monomer contacts, $n_c$, are shown. The low-temperature 
peak in the specific
heat coincides with a jump of $n_c$: at lower temperatures
the protein is swollen, and maximizes the number of water-H contacts, in
agreement
with cold denaturation. 
The number of contacts, $n_c$, begins decreasing coinciding
with the high-temperature peak of the specific heat, that therefore coincides 
with the usual 
warm denaturation phenomenon. 
Between $T_c$ and $T_w$ there is
a region where the most probable conformation is the one represented in the
inset of Fig.\ref{Fig2}: as it can be seen, it is compact with a hydrophobic 
core, out of reach for water (we also checked that this {\it native}
state is unique,
in that its Boltzmann weight is the largest above $T_c$). 
We have analyzed the behavior
of different protein lengths and of different sequences, and we have 
always found the same
qualitative behavior of $C_v$ and $n_c$.
Our model is therefore able to describe, within a single framework, 
both cold and warm denaturation. Moreover, it shows a native state
with a mostly hydrophobic core.

Although the ratio between $T_c$ and $T_w$ 
in Fig.\ref{Fig2} is unphysical, using the full MLG model it is possible to 
come closer to real values: the price to be paid is the larger number
of parameters to adjust. In this Letter we address the physical principles
responsible for the thermodynamic behavior of proteins on a broad range of
temperatures: we believe that the differences between the bimodal model and the 
MLG model (and other possible more refined models) govern the details of the
behavior more than the essential features.

We next compare the free energy, enthalpy and entropy variations of
folding of our model with those from the literature\cite{MPG90,Privalov1}.
Indeed, such a comparison is a difficult one, since it is hard
to define what a denatured state is in our theoretical calculations.
Therefore, as a simple approximation, 
we consider as denaturate those conformations with at most 
$4$ monomer-monomer contacts (a polymer of $17$ monomers over a square lattice
has at most $9$ monomer-monomer contacts). 
The native state
has $8$ monomer-monomer contacts.

\begin{figure}
\centerline{\psfig{file=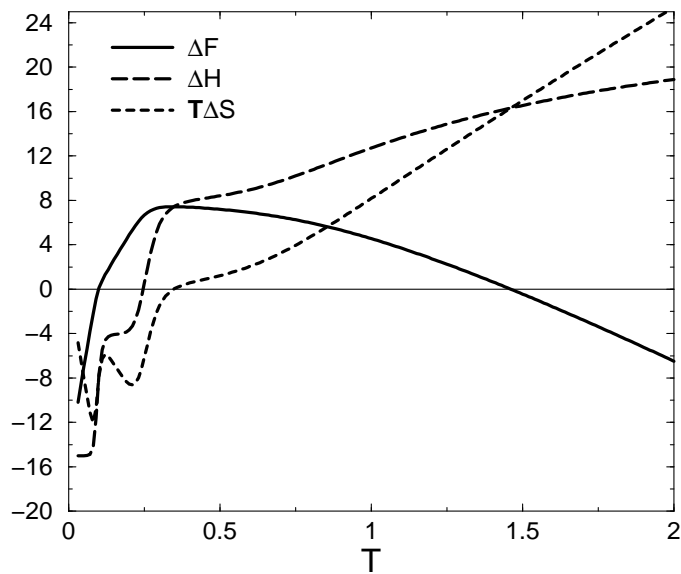,width=9.0cm}}
\caption{Free energy, enthalpy and entropy (times $T$) 
differences between denatured conformations and the native one (shown in the
inset of Fig.2), for the same parameter values as in 
Fig.2. Since $T \Delta S$ grows linearly at high temperatures,
$\Delta S$ saturates.}
\label{Fig3}
\end{figure}

In Fig.\ref{Fig3} we show $F_{Denaturate}-F_{Native} = \Delta F^D_N$,
$\Delta H^D_N$ and $T \Delta S^D_N$. They coincide qualitatively
with the ones from experiments\cite{MPG90,Privalov1}. 
We point out the presence of two temperatures
below and above which $\Delta F^D_N < 0$: the denatured state of our model
protein is more stable than the native state. 
Between these two temperatures, instead,
$\Delta F^D_N > 0$, and the native state is the most stable. 
In the same temperature range where $\Delta F^D_N > 0$, $\Delta H^D_N$ and 
$T \Delta S^D_N$ have a strong temperature dependence: they even change sign, a
signature of the rich physics behind the water-protein system.
At high temperatures we find that both $\Delta H^D_N$ and $\Delta S^D_N$
saturate ($T \Delta S$ grows linearly, therefore $\Delta S$ saturates), 
as experimentally observed\cite{Privalov1}.
Some particular care should be paid to the low temperature behavior of
$\Delta H^D_N$ and $T \Delta S^D_N$. Indeed, $\Delta H^D_N$ goes to a constant
value, which is consistent with a lower bound for the energies, and 
$T \Delta S^D_N$ tends to $0$ with $T$. Experiments should be made
below $T_c$ to assess such a behavior (although a recent model suggests 
such scenario~\cite{SHD98}).
We find therefore that our model reproduces qualitatively
the known calorimetric data of protein denaturation over a broad range of
temperatures. 
    
\begin{figure}
\centerline{\psfig{file=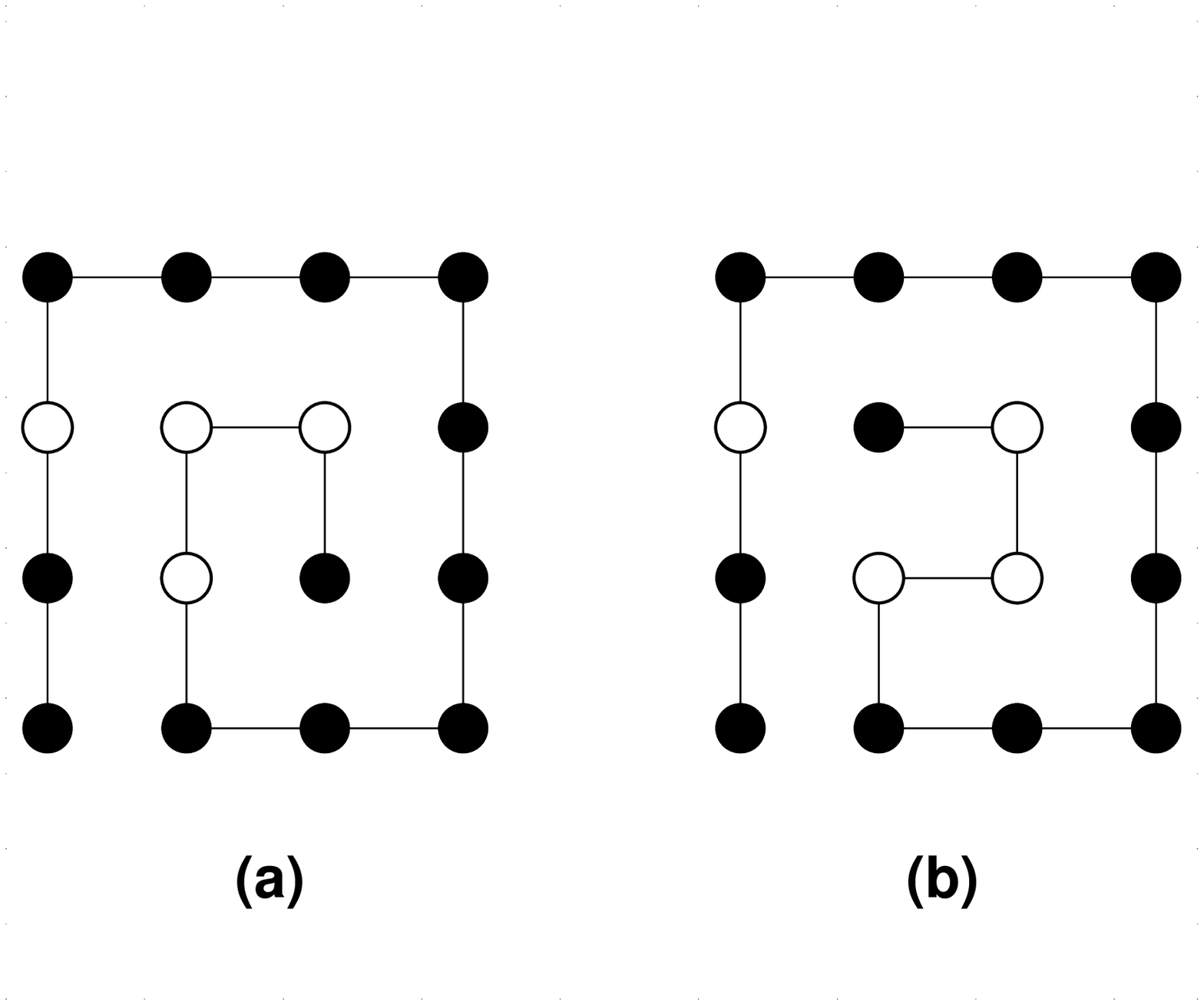,width=8.0cm,angle=270}}
\caption{Two different conformations of the same sequence
differing only for a reorganization of the core amino-acids.}
\label{Fig4}
\end{figure}

The hydrophobic effect
is often modeled through attractive  effective $HH$ interactions. 
Within our framework, we consider a system of two $H$ amino-acids in solution and we compare
the partition function of the system when the two amino-acids
are in contact, $Z_c$, with the one when the two
amino-acids have no mutual contacts $Z_0$. The effective attractive
interaction is defined as $\epsilon = T \ln({Z_c}/{Z_0})$ ($\epsilon$
is positive if attractive, with this definition).
The $T\to\infty$ limit is 
\begin{equation}
\epsilon(T\to\infty) = 2K - \frac{2}{q}(K+J)
\end{equation}
and is attractive for large values of $q$: it is the usual 
hydrophobic effective interaction. Yet, the $T=0$ limit
is $\epsilon=-2J$, repulsive. A meaningful effective interaction 
should at least include such a temperature dependence.
Actually, the strong temperature dependence of $\epsilon$ is not the
only limitation to a definition of an attractive $HH$ interaction.
Indeed, such an interaction can be meaningful only for amino-acids 
surrounded by water molecules, but it cannot be defined in the core
of proteins, where water is absent. As a consequence, in the
absence of some true interactions between amino-acids, the hydrophobic 
interaction alone is not able to favor thermodynamically
the native state against different compact states obtained by reordering
only the core of the protein. As an example, the two conformations
in Fig.\ref{Fig4}, corresponding to the sequence $PPHPPPPPPPPPHHHP$
have the same probability to occur in our model, since they hide and expose to
water the same number of H amino-acids.

Therefore this model suggests that it is improper to define 
interactions of hydrophobic origin inside proteins, and that 
the detailed structure of the cores of proteins should be stabilized
by other mechanisms. Indeed, in the biochemistry literature the debate
is still strong whether the hydrophobic interaction alone
is able to enforce the full native state of proteins or other interactions 
should also be taken into account\cite{Skolnik}. 
Effective interactions can be safely defined whenever they substitute 
some non-changing environment. When a protein is folding, its amino-acids
find instead an {\it ever changing} environment that depends on water and 
on the other amino-acids. Even the 
reliability of two-body effective interactions vs. many-body ones is an open 
issue still to be settled.
It is therefore intrinsically difficult to define 
effective potentials of some general validity between amino-acids:
our model points out such a problem for hydrophobic interactions.

In conclusion, we have introduced a model of proteins in water
that is able to reproduce the known features of proteins 
(namely, cold destabilisation and warm denaturation, 
a native state with a mostly hydrophobic core, 
and the correct free energy, enthalpy and entropy
of folding). 
We also checked our results for different protein lengths,
sequences, parameter values and even implementing the full MLG model
for the description of water. Although some details may change,
the overall behavior is consistent and robust. Moreover, lattice models
are intended to be only qualitatively instructive, whereas a quantitative
description can be given only by more detailed off-lattice models.

\begin{thebibliography}{99}

\bibitem{LD89+} K.F. Lau and K.A. Dill, Macromolecules {\bf 22}, 3986 (1989);
H.S. Chan and K.A. Dill, Physics Today {\bf 46}, 24 (1993).

\bibitem{Vanderzande} C. Vanderzande, {\it Lattice Models of Polymers} 
(Cambridge University Press, Cambridge 1998).

\bibitem{Kauzmann59} W. Kauzmann, Adv. Protein Chem. {\bf 14}, 1 (1959).

\bibitem{MPG90} K.P. Murphy, P.L. Privalov and S.J. Gill, Science {\bf 247},
559 (1990).

\bibitem{Privalov1}  P.L. Privalov and S.J. Gill, Adv. Protein Chem. {\bf 39}, 
191 (1988); G.I. Makhatadze and P.L. Privalov, J. Mol. Biol. {\bf 213}, 375 
(1990); G.I. Makhatadze and P.L. Privalov, J. Mol. Biol. {\bf 232}, 639 (1993);
P.L. Privalov and G.I. Makhatadze, J. Mol. Biol. {\bf 232}, 660 (1993);
G.I. Makhatadze and P.L. Privalov, Adv. Protein Chem. {\bf 47}, 307 (1995) and
references therein.

\bibitem{Privalov2} see {\it e.g.}
N.C. Pace and C. Tanford, Biochemistry {\bf 7}, 198 (1968);
P.L. Privalov, Yu. V. Griko, S. Yu. Venyaminov and V.P. Kutyshenko,
J. Mol. Biol. {\bf 190}, 487 (1986);
Y.V. Griko and P.L. Privalov, Biochemistry {\bf 31}, 8810 (1992);
J. Zhang, X. Peng, A. Jonas and J. Jonas, Biochemistry {\bf 34}, 8631 (1995);
E.M. Nicholson and J.M. Scholtz, Biochemistry {\bf 35}, 11369 (1996); G.
Graziano, F. Catanzano, A. Riccio and G. Barone, J. Biochem. {\bf 122}, 395
(1997);  
for a general review see 
P.L. Privalov, CRC Crit. Rev. Biochem. Mol. Biol. 
{\bf 25}, 181 (1990), and references therein.

\bibitem{FE45+} H.S. Franks and M.W. Evans, J. Chem. Phys. {\bf 13}, 507 (1945);
see also T.E. Creighton, {\it Proteins. Structures and Molecular
Properties} (W.H. Freeman \& Company, New York 1993), 157-161, 
and references therein.

\bibitem{Muller90} N. Muller, Acc. Chem. Res. {\bf 23}, 23 (1990);

\bibitem{LG96} B. Lee and G. Graziano, J. Am. Chem. Soc. {\bf 22}, 5163 (1996);

\bibitem{SHD98} K.A.T. Silverstein, A.D.J. Haymet and K.A. Dill,
J. Am. Chem. Soc. {\bf 120}, 3166 (1998).

\bibitem{SHD99} K.A.T. Silverstein, A.D.J. Haymet and K.A. Dill,
J. Chem. Phys. {\bf 111}, 8000 (1999).

\bibitem{Skolnik} Many methods to predict the native state of proteins 
have been recently developed, based on residue-residue interactions of a 
statistical origin, without any specific assumption on the "real", 
microscopic, nature of such potentials. Their derivation from first principles,
when possible, is still an open issue; see e.g. S. Miyazawa and R.L. Jernigan,
J. Mol. Biol. {bf 256}, 623 (1996); J. Skolnick, L. Jaroszewski, A. Kolinski and
A. Godzik, Protein Sci. {\bf 6}, 676 (1997).


\end{thebibliography}
\end{document}